\documentclass[epsf,twocolumn,preprintnumbers]{revtex4}
\usepackage{graphics}
\usepackage{graphicx}
\usepackage{dcolumn} 
\usepackage{bm}
\usepackage{color}
\usepackage{epsfig}
\usepackage{comment}
\usepackage[normalem]{ulem}
\newcommand{\bco}{Ba$_2$CuO$_3$}
\newcommand{\bcox}{Ba$_2$CuO$_{3+\delta}$}
\newcommand{\sco}{Sr$_2$CuO$_3$}
\newcommand{\scox}{Sr$_2$CuO$_{3+\delta}$}
\newcommand{\beq}{\begin{eqnarray}}
\newcommand{\eeq}{\end{eqnarray}}

\begin{document}
\title{Two-band Conduction and Nesting Instabilities in \\
  Superconducting Ba$_2$CuO$_{3+\delta}$: a First Principles Study
}
\author{Hyo-Sun Jin$^1$}
\author{Warren E. Pickett$^{2}$}
\email{wepickett@ucdavis.edu} 
\author{Kwan-Woo Lee$^{1,3}$}
\email{mckwan@korea.ac.kr} 
\affiliation{
 $^1$Division of Display and Semiconductor Physics, Korea University, Sejong 30019, Korea\\
 $^2$Department of Physics, University of California, Davis, CA 95616, USA\\
 $^3$Department of Applied Physics, Graduate School, Korea University, Sejong 30019, Korea
}
\date{\today}

\begin{abstract}
First principles investigations of the high temperature superconducting system
Ba$_2$CuO$_{3+\delta}$, recently discovered at $\delta\approx0.2$ at $T_c=70$ K,
are applied to demonstrate the effects of oxygen ordering on the electronic
and magnetic properties. The observed `highly over-doped' superconducting phase
displays stretched Cu-planar oxygen O$_{\rm P}$ distances and anomalously
shortened Cu-apical O$_{\rm A}$ separations
compared with other cuprates.
The stoichiometric system $\delta=0$, with its strongly one-dimensional (1D)
Cu-O$_{\rm P}$ chain structure,  when nonmagnetic shows 1D Fermi surfaces
that lead, within density functional theory,
to antiferromagnetic Cu-O$_{\rm P}$ chains (a spin-Peierls instability).
Accounting for 1D fluctuations and
small interchain coupling according to the theory of Schulz indicates this
system, like Sr$_2$CuO$_3$,
is near the 1D Luttinger-liquid quantum critical phase.
The unusual Cu-O bond lengths {\it per se} have limited effects on other properties
for $\delta$=0.
We find that a `doubled bilayer' structure of alternating Cu-O$_{\rm P}$ chains
and wide rung Cu$_3$O$_4$ ladders is the energetically preferred one
of three possibilities where
the additional oxygen ions bridge Cu-O$_{\rm P}$ chains in the superconducting phase $\delta=1/4$.
Nominal formal valences of the three Cu sites are discussed. The six-fold
(octahedral) site is the most highly oxidized, accepting somewhat more holes in the
$d_{z^2}$ orbital than in the $d_{x^2-y^2}$ orbital. The implication is that two-band
physics is involved in the pairing mechanism and the superconducting carriers.
The Fermi surfaces of this metallic bilayer structure show both 1D and 2D
strong (incipient) nesting
instabilities, possibly accounting for the lack of clean single-phase samples
based on this structure and suggesting importance for the pairing mechanism.
\end{abstract}

\maketitle

\section{Introduction}
In superconducting cuprates, the CuO$_2$ square lattice has been widely considered
a crucial ingredient in producing high $T_c$ due to the profuse number of
examples \cite{wep1988}.
However, some more structurally complex cuprates include
Cu-O chains. An example of one class is the `telephone number compounds'
exemplified by Sr$_{14}$Cu$_{24}$O$_{41}$ with a mean copper valence of
Cu$^{2.25+}$, which is in the same doping regime as the 
superconducting square lattice
cuprates. For issues and references, see {\it e.g.}
Schmidt {\it et al.} \cite{schmidt2007} and Armstrong {\it et al.} \cite{armstr1991}.
The Cu-O chains in these compounds were found
to be plagued (or, depending on viewpoint, blessed) with inhomogeneities,
likely related to one dimensionality and spin dimerization. What is
necessary, versus what is peripheral,
for high T$_c$ superconductivity in cuprates remains under discussion.

Over 30 years ago \cite{abb1988,armstr1991,zhang1990,forget1997,cwchu1994}
\bcox~ was synthesized in the range of $\delta=0-0.4$
and refined structurally based on the K$_2$NiF$_4$ structure
of the $\delta=1$ phase, {\it i.e.} a Ba$_2$CuO$_{4-\alpha}$ picture.
At that time no superconductivity in the system was observed.
Recently, Li {\it et al.} successfully obtained a bulk superconducting
sample with a high critical temperature $T_c\approx70$ K
at $\delta\approx0.2$ \cite{pnas2019,jsnm2020}. Their synthesis technique
used  a very high pressure of 18 GPa and high temperature of 1000 C$^\circ$.
Through x-ray absorption spectroscopy (XAS) and resonant inelastic
x-ray scattering (RIXS) measurements
on a superconducting powder sample \cite{sala2021},
Fumagalli {\it et al.} inferred two inequivalent Cu ions and an
in-plane nearest neighbor superexchange $J\sim 0.12-0.18$ eV.
Theoretical studies \cite{scalapino2019,lieb2020,jinang2021} based on two-band models
proposed unconventional superconductivity of either $d$-wave or $s_\pm$-wave pairing.

Compared with other superconducting cuprates,
this  extraordinarily heavily doped superconductor
(Cu$^{p}$, $p\sim 2.4-2.6$) 
also shows notable distinctions  in structure.
It has a substantially stretched Cu-planar O$_{\rm P}$ distance $d_{\rm O_P}$
of 2.00 \AA~(normally around 1.92 \AA),
which may partially be due to the large Ba$^{2+}$ ionic radius,
and concomitantly occurring, yet very significantly compressed
Cu-apical O$_{\rm A}$ distance $d_{\rm O_A}$ of 1.86 \AA~
(compare 2.4 \AA~ in La$_2$CuO$_4$).
These values are substantially different from
the isovalent (and isostructural) superconductor \scox~ with
up to $T_c$=95 K \cite{hiroi1993,liu2006,liu2014,pnas2020},
where $d_{\rm O_P}$ and $d_{\rm O_A}$ are nearly identical at 1.95 \AA.
Additionally, in \bcox~  reduction of oxygen concentration leads to
a high fraction of vacancies in the  O$_{\rm P}$ sites,
not the O$_{\rm A}$ sites \cite{abb1988,zhang1990,pnas2019},
whereas the superconducting system \scox~ has vacancies in both
O$_{\rm P}$ and O$_{\rm A}$ sites \cite{pnas2020}.

Several of these observations have been confirmed by theoretical calculations \cite{ladder2019}.
Large oxygen O$_P$ vacancy concentrations
break up the CuO$_2$ square lattices,
leading to a one-dimensional (1D) Cu-O$_{\rm P}$ chain
lattice \cite{abb1988,zhang1990,c.jin2015}.
Stoichiometric \bco, {\it i.e.} the $\delta=0$ phase, would have a
pure 1D Cu-O$_{\rm P}$ chain lattice,
but that stoichiometry has not been well investigated yet.
The chain lattice in the sister compound \sco~ has been intensively investigated
as an ideal 1D Heisenberg $S=\frac{1}{2}$ antiferromagnetic (AFM) system
for over 20 years \cite{ami1995,thurber2001,schlappa2012,schlappa2018,serge2020}.

In this paper, we will investigate the $\delta=0$ and $\delta=1/4$ phases through {\it ab initio} calculations.
The latter one approximates the superconducting phase around $\delta\approx0.2$.
For the $\delta=0$ phase of Cu$^{2+}$, compared with those of \sco,
the half-filled $d_{X^2-Y^2}$ orbital (notation to be established later)
leads to half the interchain hopping strength
but nearly identical intrachain hopping strength,
suggesting this compound provides a more nearly ideal 1D Heisenberg
$S=\frac{1}{2}$ AFM system, as will be discussed.
Considering AFM order on the chain, the system is
gapped by a 1D spin-Peierls instability at $q=2k_F$. 
In contrast to the previous suggestion that the substantially compressed $d_{\rm O_A}$
leads to inverted crystal field splittings of the $e_g$ orbitals \cite{pnas2019},
our calculations indicate that this system is not greatly affected by the
unusual Cu-O separations.
The energy gap, however, is substantially changed by variation of
the Cu-O$_{\rm P}$ separation and by O$_{\rm A}$ phonon modes,
likely related to the spin-Peierls instability.

The structures and computational methods we have used are described in
Sec. II. In Section III, a description of the nonmagnetic $\delta$=$0$
electronic structure is provided, followed by that of the magnetic
(AFM) electronic structure and magnetic characteristics. Effects on
the band structure of high symmetry O displacements are presented
in Sec.~III.D. 
Based on the inferred vacancies at the planar oxygen sites in \bcox,
several lattices \cite{lieb2020,ladder2019,ladder2020,brick} with
ordered planar oxygens have been suggested
to approximate the superconducting phase.
Among them, we have investigated  in Sec. IV  
structures with brickwall \cite{brick}
and two-type ladder plus chain lattices \cite{ladder2019,ladder2020},
which have Cu ions in up to three types of coordination.
At $\delta=1/4$ doping,
the bilayer structure of a ladder layer plus a chain layer is found to be
the ground state among the three structures.
This structure contains all three Cu coordinations, displaying both
(roughly) half-filled and quarter-filled Cu orbitals.
Copper oxidation states, $e_g$ orbital characteristics, and characteristics
of Cu coordination (four, five, and six) are compared to identify
distinctive aspects of the favored bilayer structure.

In Sec. V the fermiology of the energetically favored bilayer structure is
presented in more detail, to inform later experimental data. Section VI
presents a discussion of our results, including some relationships to the
long standing studies of spin-half chains. A summary is provided in
Sec. VII. Some additional results are presented in the Appendix.

\begin{figure}[tbp]
{\resizebox{8cm}{6cm}{\includegraphics{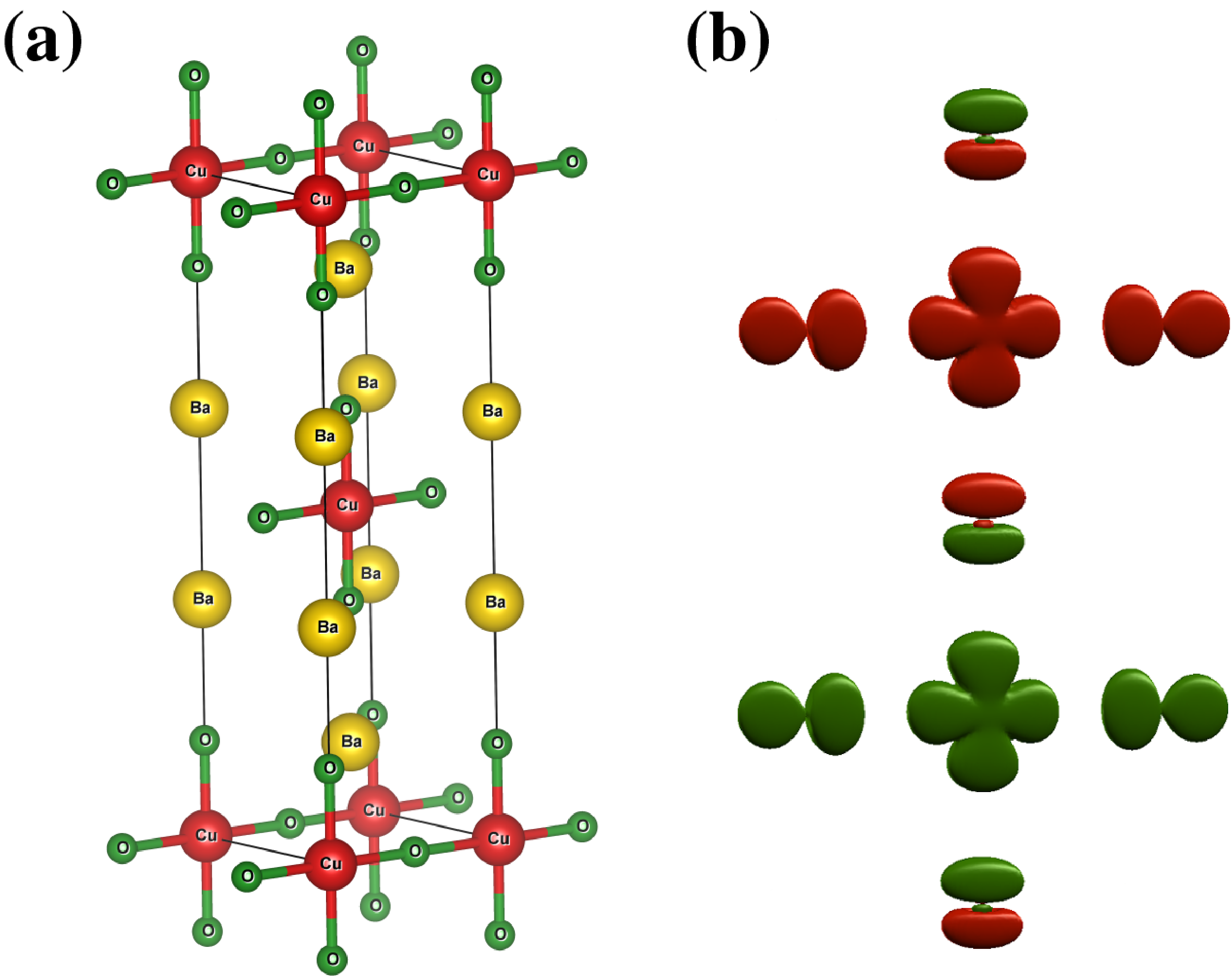}}}
\caption{(a) Idealized crystal structure of \bco, based on the 1D Cu-O$_{\rm P}$ chain.
The apical oxygens O$_{\rm A}$ and Ba ions sit at nearly identical heights.
The Cu-O$_{\rm P}$ chains are well separated by the insulating BaO$_{\rm A}$ layer, 
by 6.57 \AA~ along the $\hat{c}$-direction.
(b) Spin density plot (GGA) of the  \bco~ AFM state, 
showing the strong $pd\sigma$ hybridization.
The isovalue is 0.015 e/\AA$^3$. Each spin character is described 
by the different colors.
}
\label{str}
\end{figure}

\begin{figure*}[htbp]
{\scalebox{0.18}{\includegraphics{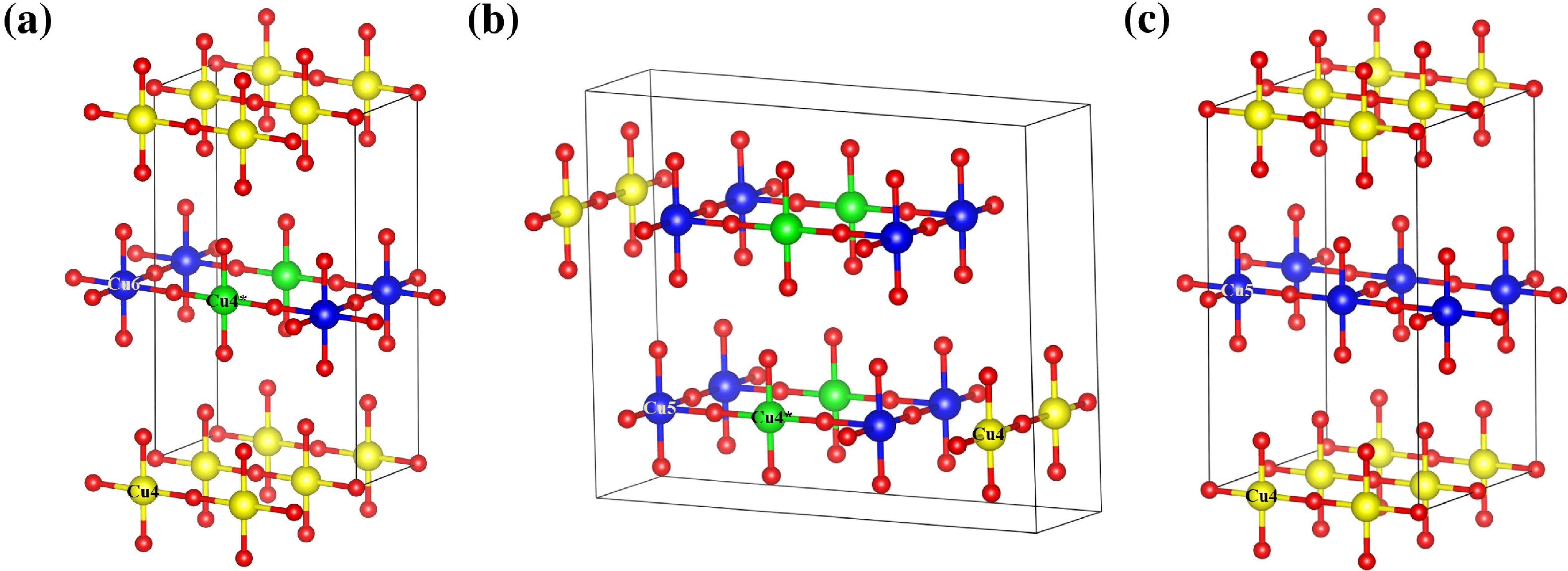}}}
\caption{Crystal structures, investigated here,  for the $\delta=1/4$ phase.
For simplicity, only Cu (larger spheres) and O (smaller, red spheres) ions are shown.
 Both the bilayer (a) and brickwall (c) structures have 1D Cu-O chains 
 extending horizontally in the $z=0$ layer,
 whereas the monolayer (b) phase  has the same structure in both the $z=0$ 
 and the $z=\frac{c}{2}$ layers. Each structure is derived from the \bco~chain
 structure, by (i) inserting O ions linking the chains, and (ii) moving the planar
 O on alternating chains to complete the ladder rung, or the long edge of the
 brick.
 We used supercells of  $2\times1\times1$ for (a), $1\times4\times1$ for (b), 
 and $2\times2\times1$ for (c).
 Here, the $\hat{a}$-direction is always chosen along the Cu4-O$_{\rm P}$ chain line.
}
\label{str_sc}
\end{figure*}

\section{Methods and Structures} 
\subsection{Methods of calculation}
We initially carried out first principles calculations using the generalized 
gradient approximation (GGA) \cite{gga} for exchange and correlation,
which is implemented in the all-electron full-potential code {\sc wien2k} \cite{wien2k}.
The results for the AFM phase at $\delta=0$
were confirmed by another all-electron full-potential code {\sc fplo-18} \cite{fplo}
(see the Supplemental  Material  (SM) \cite{sm}).

In {\sc wien2k}, the basis size was determined by $R_{mt}K_{max}=7$
with augmented-plane-wave sphere radii $R_{mt}$ (in $a.u.$): Cu, 1.71;
O, 1.47; and 2.50 for Ba.
For {\sc fplo} the default orbitals were used.
For both codes, the Brillouin zone was sampled by a dense mesh
containing up to 8,000 $k$-points 
to check the energetics and fine band structures (specifically, small
energy gaps) carefully.

Structural optimization for the doped cases was carried out with the {\sc VASP} code \cite{vasp}, 
using an energy cutoff of 600 eV. This code is efficient for the supercells, which 
contain up to 50 atoms. The lattice parameters were optimized, while the
internal structural parameters were relaxed until all forces were
less than 0.01 eV/\AA. (See SM for further information \cite{sm}.) 

Correlation effects of the electronic properties are always a consideration in
cuprates. The most used procedure is to include Hubbard $U$ and Hund's exchange
$J_H$ parameters in the DFT+U procedure \cite{erik}. We have done this for the
insulating $\delta=0$ compound. For the metallic and superconducting phase $\delta\sim 1/4$,
calculations neglecting $U$ give more realistic (sometimes excellent \cite{wep1992})
Fermi surfaces and energetics while excitations, which we do not address here,
require dynamical calculations including $U$ \cite{dmft}. 

There is no single
accepted method for calculating $U$, with various proposals depending on
procedure, on the choice of `atomic orbital' (which might even be a Wannier
orbital), and on the choice of treating the dynamical correlations. We discuss
the commonly employed single band expression for nearest neighbor exchange
coupling of spins $J=4t^2/U$ to provide a guide to the magnitude of magnetic
coupling, and use $U$=7 eV in quantities we provide, unless otherwise
stated. For a range of values around this choice of $U$ (which has been 
calculated \cite{SashaL1995,aza1991,ku2002} and applied with realistic results
several times before) physical properties typically vary
modestly with the choice of $U$.

\subsection{Structures for $\delta=0$ and $\delta=1/4$}
The available structural information is ambiguous, and it seems that stoichiometric
and ordered \bco~(Cu$^{2+}$) has not been achieved. Powder x-ray diffraction
data on the $T_c$=70 K multi-phase sample was refined within the $I4/mmm$ space
group (as for La$_2$CuO$_4$), obtaining \cite{pnas2019} $a=4.00$ \AA, 
$c=12.94$ \AA, with 40\% vacancies randomly placed on the {\it planar} O sites,
obtaining the Ba$_2$CuO$_{3.2}$ composition. The in-plane lattice constant is
unexpectedly large compared  to most layered cuprates, and an O$_{\rm A}$-Cu
separation of 1.86 \AA~ was emphasized as being exceptionally small, viz. 
2.4 \AA~ in La$_2$CuO$_4$. A cluster
expansion study \cite{ladder2020} of compositions around $\delta\sim 0.2$
found that ordered supercells containing Cu-O$_{\rm P}$ chains and ladders are 
favored up to 900 K. Our choices are guided by the chain structure
of isovalent Sr$_2$CuO$_3$ for $\delta$=0, and by the simplest choices of
added in-plane oxygen atoms (described below) for $\delta$=1/4. These choices
allow the assessment of differences due to coordination (there are four-,
five-, and six-fold Cu ions), which seems to be a more useful viewpoint
than crystal field splitting.

{\it $\delta=0$: \bco.}
Our first calculations are based on the lattice parameters of $a=b=4.003$ \AA~
and $c=12.942$ \AA~\cite{pnas2019},
recently refined in a superconducting sample at room temperature as noted above.
These values are close to the early reported
values \cite{abb1988,zhang1990,c.jin2015} also on multi-phase samples.
For \bco~ with 50\% ordered vacancies on the planar oxygen site
(space group: $Immm$, No. 71),
shown in Fig. \ref{str} (a), the Cu and O$_{\rm P}$ ions sit at $2a$ (0,0,0)
and $2d$ (0,$\frac{1}{2}$,0) sites,  respectively.
The Ba and O$_{\rm A}$ ions lie on $4i$ (0,0,$\xi$) sites
with internal parameters $\xi_{\rm Ba}=0.1437$ and $\xi_{\rm O_A}=0.1438$,
at effectively identical heights.
As discussed below, calculations for \bco~ show a substantial 
tetragonal ($a\not=b$) distortion,
reflecting the chain structure.
This difference is consistent with the experimental observations in \scox~
that show a rapidly reduced difference in the planar lattice parameters,
as oxygen concentration in \scox~is increased from $\delta=0$ \cite{pnas2020}.

We began our investigations from the $\delta=0$ Cu$^{2+}$ phase that 
serves as the closest stoichiometric parent compound of the \bcox. We
use the chain structure originally proposed by Armstrong {\it et al.} \cite{armstr1991}. 
In contrast to the above mentioned $I4/mmm$ optimization of the 70 K sample,
other data on \bcox~  has indicated a difference of in-plane lattice parameters,
less than  1\% (0.04 \AA) in a high temperature, possibly strongly site 
disordered, sample \cite{zhang1990,forget1997,c.jin2015}
but a large 0.3 \AA~ difference in a low temperature sample\cite{abb1988}.
This tetragonal distortion is consistent with ordering of O vacancies, likely in
a 1D Cu-O$_{\rm P}$ chain (sub)structure.
Due to the differing reports, for the $\delta=0$ phase we fully optimized 
all structural parameters in the orthorhombic space group $Immm$.

{\it $\delta=\frac{1}{4}$: Ba$_2$CuO$_{3.25}$}.
To describe models of ordered planar oxygens in the superconducting doped regime,
we investigated three structures of doubled or quadrupled
supercells (as necessary) having two or three Cu sites, 
pictured in Fig. \ref{str_sc} \cite{ladder2019,ladder2020,brick}.
All structures are derived from the \bco~layers of 
Cu-O$_{\rm P}$ chains in which each Cu
has an apical O above and below. The structures differ in where the additional
planar O, necessarily bridging chains, per four Cu sites is distributed.
The additional ions produce five-fold coordinated Cu sites 
(denoted Cu5) along the chains,
octahedral six-fold (Cu6) sites in the bilayer structure,  
and new four-fold sites (Cu4$^{\ast}$) on the rungs in addition to 
the Cu-O$_{\rm P}$ chain site (Cu4), which is the only site for $\delta$=0.
We describe the structures specifically because the coordination 
has strong effects
on the Cu $e_g$ orbitals.

\noindent 
{\it Bilayer structure.} 
Figure \ref{str_sc}(a) illustrates the oxygen
ladder ordering, which consists of chains bridged by O-Cu-O units.
In this bilayer case the $z=\frac{c}{2}$ and  $z=0$ layers consist
of the Cu-O ladders and \bco-type chain lattices, respectively. 
In this structure two bridging O ions are inserted in alternating layers.
The $z=\frac{c}{2}$ layers can be viewed as connected ladders. 
There are one six-fold coordinated Cu ion (Cu6) and three four-fold 
Cu ions, one along each
chain (Cu4) and one along the rungs (Cu4$^{\ast}$).

\noindent 
{\it Monolayer structure.} 
In this structure the two layers are identical (hence {\it `monolayer'}),
but displaced by a body-centering operation. The additional O ions (with
respect to \bco) are shown in Fig.~\ref{str_sc}(b) to lead to a chain and a ladder. 
There are two 
five-fold coordinated Cu5 site, and two distinct four-fold sites, Cu4
along the chains and Cu4$^{\ast}$ forming the rung of the ladder.

\noindent 
{\it Brickwall structure.} 
This structure \cite{brick} consists of one layer of
Cu-O chains (Cu4 sites), two per primitive supercell. The other
layer has a brickwall ordering of oxygen vacancies,
as illustrated in Fig. \ref{str_sc}(c). 
Planar oxygen vacancies occur every two squares along one in-plane direction,
being staggered by ($a,a,0)$ in the other direction to give the brickwall
motif. Each copper site in this layer is bonded to two apical O ions and
three planar O ions, thus being five-fold coordinated Cu5 sites.

\begin{figure}[tbp]
{\resizebox{8cm}{5cm}{\includegraphics{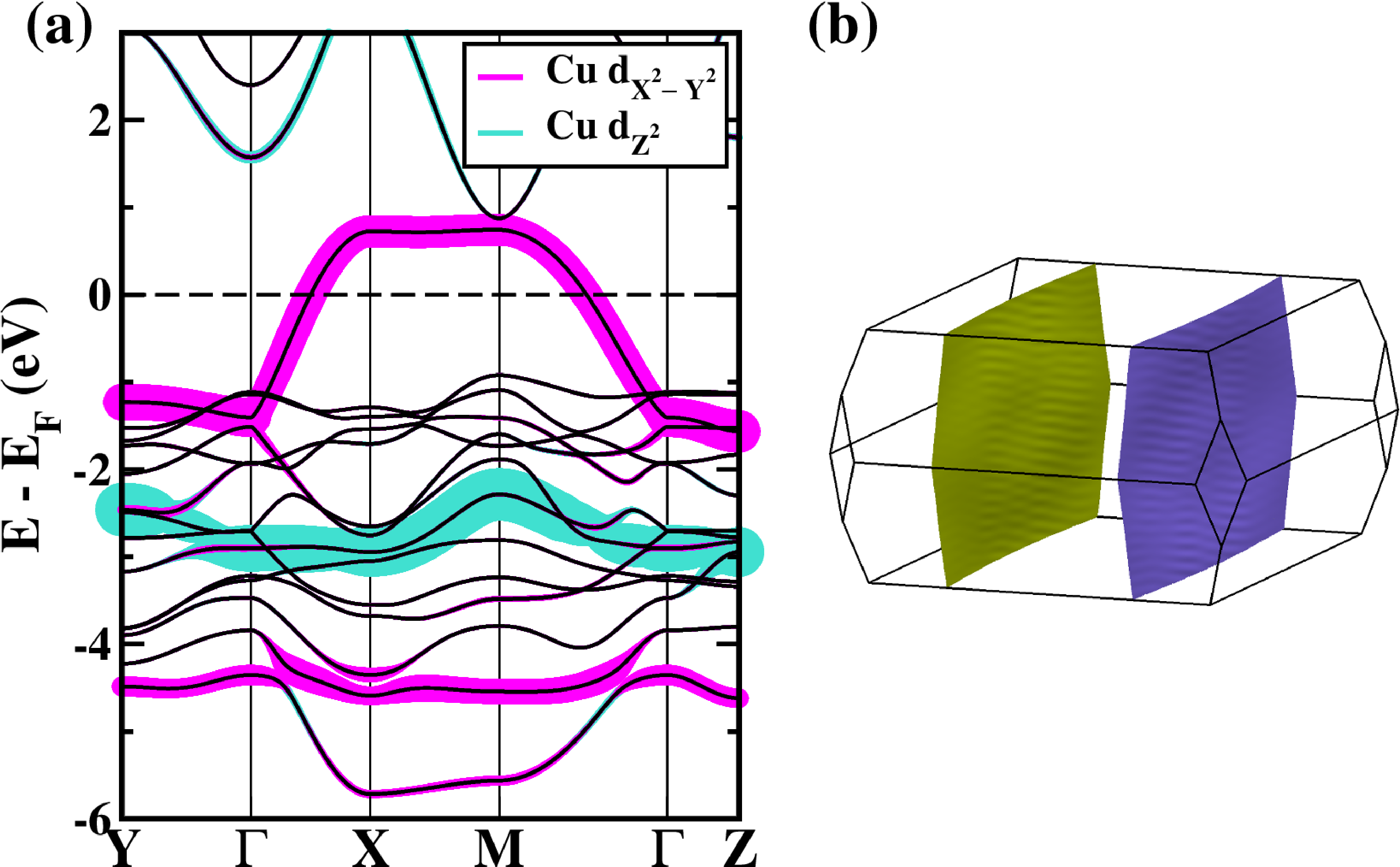}}}
\caption{ (a) GGA nonmagnetic band structure for the optimized structure,
with fatband representation of the Cu $d_{X^2-Y^2}$ and $d_{Z^2}$ characters
in the range of --6 eV to 3 eV, containing the Cu $3d$ and O $2p$ orbitals \bco. 
The $X$ and $Y$ points are chosen as the zone boundaries 
parallel and perpendicular to the Cu-O$_{\rm P}$ chain, respectively.
(b) The corresponding Fermi surfaces, 
showing very strong 1D character arising from the Cu-O$_{\rm P}$ chain band.
} 
\label{nmband}
\end{figure}

\section{Results for parent $\delta=0$ phase}

\subsection{Structural items}
Structural differences between nonmagnetic (NM) and AFM states 
were on the order of 10$^{-2}$\AA.
The optimized NM lattice parameters are $a=4.144$, 
$b=3.808$, $c=13.142$ (in units of \AA),
with the internal parameters of $\xi_{\rm Ba}=0.1459$, 
$\xi_{\rm O_A}=0.1481$.
The corresponding Cu-O$_{\rm P}$ and Cu-O$_{\rm A}$  distances 
are 2.07 \AA~ and 1.95 \AA~respectively, a difference of $\pm$6\%.
The lattice parameters are consistent with the early experimental 
low temperature values \cite{abb1988}.
As expected, the Cu-O$_{\rm P}$ chains lie along the 
longer $\hat{a}$-direction.
Compared with the experimental value of the superconducting sample \cite{pnas2019}, 
the Cu-O distances are larger by 0.07-0.08 \AA, but the 
interchain distance is smaller by 0.2 \AA~ in the layer, very significant
differences.
For our optimized structures with the same cell volumes \cite{optimize},
the $a\not=b$ case has a lower energy by 30 meV/f.u. than for  $a=b$.
Thus for the $\delta=0$ we will focus on results for the optimized 
$a\not=b$ structure, unless specified otherwise.

In both $a=b$ and $a\not=b$ cases, 
the AFM state with alternating spin alignment along the Cu-O$_{\rm P}$ chain 
has a lower energy than in the NM state,  
by 4 meV/f.u. for the $a\not=b$ case.
This difference is remarkably small, but comparable with the 
energy gain of about 7 meV/f.u. 
from a simple Stoner picture $I_{st}M^2/4$ with Stoner $I_{st}=0.9$ 
eV \cite{erik} and the Cu moment of 0.18 $\mu_B$ (see below).
This indifference of the AFM energy gain to the $a/b$ ratio is
another property that is insensitive to the tetragonal distortion 
in the planar lattice parameters.
Our attempts to obtain a ferromagnetic state always reverted to NM states.

\begin{figure}[tbp]
{\resizebox{8cm}{8cm}{\includegraphics{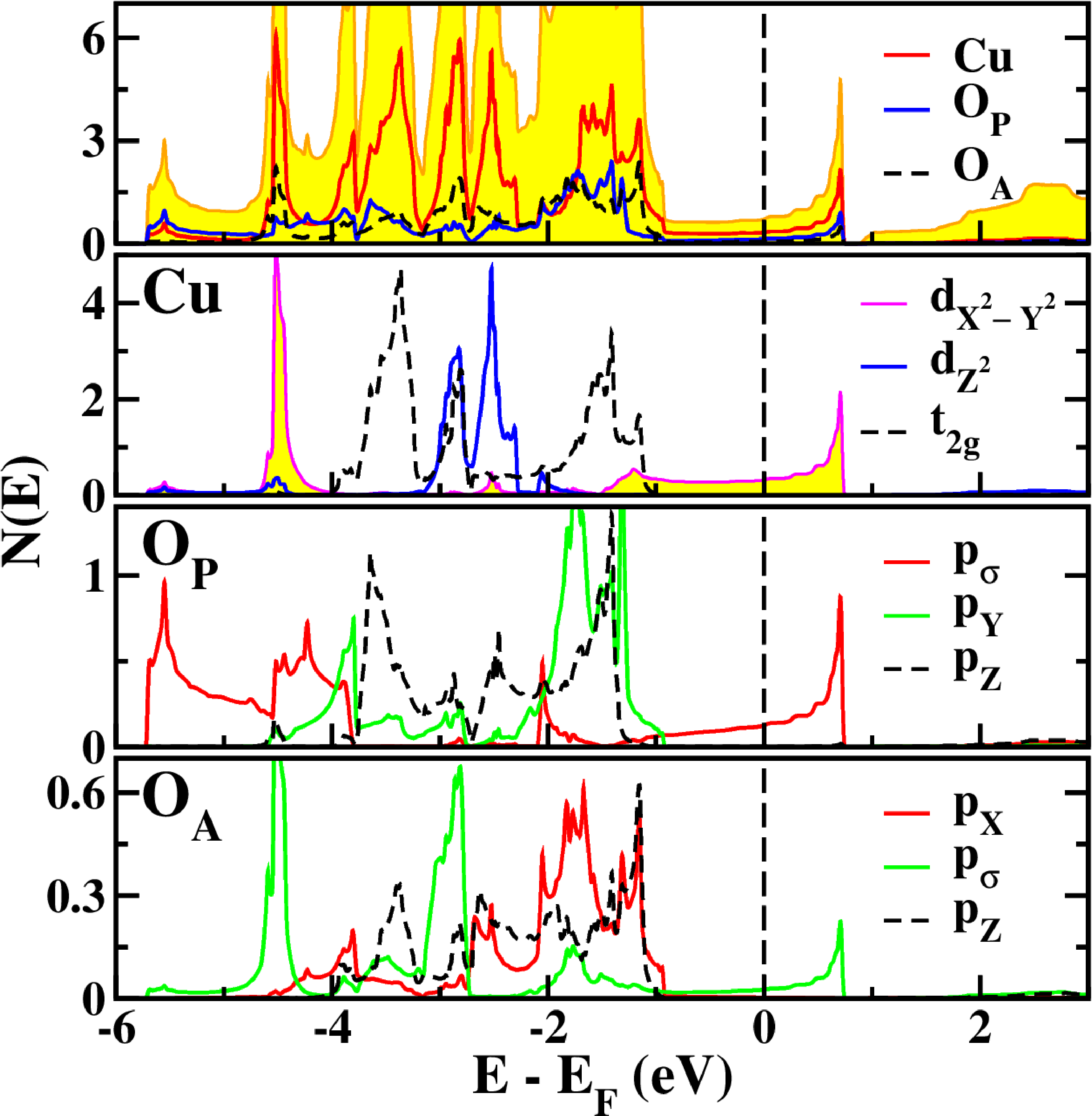}}}
\caption{ For NM  \bco,
(top) total and atom-projected densities of states,
 and (bottom) orbital-projected densities of states for each atom, within GGA.
The total Fermi level DOS $N(E_F)$ = 0.72 eV$^{-1}$ per f.u. for both spins.
The total DOS, yellowish-shaded in the top panel, varies little around $E_F$, 
reflecting the mid-band 1D character.
Due to the chain structure, the mostly filled O $p_\pi$ orbitals 
show differences between the two sites.
} 
\label{nmdos}
\end{figure}

\subsection{Nonmagnetic electronic structure}
Now we address the NM state to establish the underlying 
(spin-symmetric) electronic structure.
The bands and corresponding atom and orbital projected-densities of states (PDOSs) 
are provided in Figs. \ref{nmband}(a) and \ref{nmdos}, respectively.
These are similar to the previously reported ones in the $a=b$ case \cite{jinang2021}.
To analyze this, we focus on the CuO$_4$ unit and choose the $X$-axis 
along the Cu-O$_{\rm P}$ chain and the $Y$-axis toward the apical oxygen ${\rm O_A}$.
Consistent with the Cu$^{2+}$ $d^9$ configuration,
the $d_{X^2-Y^2}$ antibonding state with a band width of 2.3 eV 
crosses the Fermi energy $E_F$ exactly at half filling.
The less dispersive $d_{Z^2}$ orbital (perpendicular to the CuO$_4$ unit)
is occupied at --3 eV. 
The half-filled $d_{X^2-Y^2}$ band leads to a strongly 1D Fermi surface (FS), 
displayed in Fig. \ref{nmband}(b), suggesting strong 1D instabilities 
at $q=2k_F$ (see below).

Figure~\ref{nmband} suggests that a simple single band model is a reasonable
starting point for \bco. 
Fitting the $d_{X^2-Y^2}$ band leads to the following nearest neighbor 
(NN) and next NN hopping parameters: \\ 
$\bullet$ a site energy of --0.146 eV\\
$\bullet$ an intrachain hopping $t_a=0.532$ eV\\
$\bullet$ a second NN intrachain hopping $t'_a=0.083$ eV\\
$\bullet$ an interchain hopping $t_b=0.044$ eV\\
$\bullet$ a second neighbor hopping $t_{ab}=0.015$ eV\\
$\bullet$ an  interlayer hopping $t_c=-0.010$ eV.\\
Keep in mind that these values correspond in principle to a Cu-centered
Wannier function with $d_{X^2-Y^2}$ symmetry and character.
Compared with our calculated hopping parameters of \sco~with the
experimental structure parameters of Ami {\it et al.} \cite{ami1995},
$t_b$ is reduced by half; changes in the
other parameters are unimportant.
This comparison indicates that \bco~is an even more nearly ideal 1D Heisenberg
AFM spin-half chain system than \sco~(see below).

A first interest in magnetic insulators such as this is
the exchange couplings between magnetic ions. These hopping parameters indicate
several exchange couplings $J_a\equiv J$, $J_z', J_b, J_{ab}$, and $J_c$.
The dominant intrachain hopping $t_a$ is conventionally used to 
estimate a corresponding superexchange strength
$J_a=4t_a^2/U$. For the value of $U$=7 eV that we have used in the
following calculations, one obtains $J_a \approx0.162$ eV ({\it i.e.}, about 1875 K).
With the value $U=8$ eV from a spectroscopic (high energy) experiment
\cite{sala2021}, the estimate reduces accordingly, to 1640 K. There have
been suggestions that $U$ as small as 5 eV might be realistic (at
least in square lattice cuprates), leading
to a considerably larger value of $J_a$.

It is useful to keep in mind the origins of the uncertainty in choosing,
or fitting to experiment, a single value of NN $J$. The bigger picture is
that the superconducting composition $\delta\sim\frac{1}{4}$ is the prime
interest, and our calculations below indicate the importance of the 
$d_{z^2}$ orbital, so a single band picture no longer holds. Even when
the additional parameters (site energies and crystal field splitting,
charge transfer energy, altered hopping amplitudes) become available,
the various exchange couplings become challenging to obtain, with 4th-order
perturbation theory expressions not being accurate \cite{eskes1993}.
There are the usual complications, that a high energy value of $U$ 
(photoelectron spectroscopy) is different \cite{biermann2003} from low energy values
(transport, spin waves), and also that $U$ values may be different
for the two $e_g$ orbitals \cite{wepUcalc1998}. Other differences in
principle also arise, such as that DFT+U codes nearly always deal
with ($U$ corrections to) atomic orbitals, whereas model Hamiltonians
conventionally neglect overlap integrals, thus assuming Wannier
functions as the basis. Our parameter values (above) provide the
starting point for more elaborate model studies of \bcox.

Focusing on the exchange parameters, a conventional practice when
there are stable localized moments is to compare the energy differences
for different alignments of the moments calculated using the appropriate
functional (usually DFT+U rather than DFT alone). AFM versus FM alignment
gives an estimate of the NN coupling (here, $J_a$). Extending to
larger supercells, other lower symmetry alignments provide more energies,
and a number of exchange constants can be fit to them. We have found that,
for the value $U$=7 eV for which we report results, no FM
can be obtained; the moment iterates to zero. Since the AFM state (and
also no FM state) was obtained even for $U$=0 (see below), the lack of
FM solution confirms that the AFM moments we do obtain are not of
primarily local character. Also, their magnitudes are smaller than
expected of spin-half moments, even accounting for hybridization with oxygen. 
A modest FM moment would result in exchange split bands, which
from Fig.~\ref{nmband}(a) would leave a metallic, not insulating, electronic
structure, uncharacteristic of single band local moments. A reliable
means of obtaining the exchange parameters would be to perform 
self-consistent linear response calculations of the spin wave
spectrum \cite{spinwaves}, which would be very useful if stoichiometric
\bco~can be synthesized and characterized. 

As mentioned above, 
effects of the tetragonal distortion of the $a$-$b$ lattice parameters on the 
electronic structures and magnetic characters are not substantial.
However, a few distinctions can be observed.
The distortion decreases $t_{a}$ by 10\% and increases $t_{b}$ 
by a factor of $\frac{3}{2}$,
reflecting the increased intrachain and reduced interchain Cu-O$_{\rm P}$ separations.
In both NM and AFM states,
a clearly distinguishable change is to increase a gap between 
the $E_F$-crossing $d_{X^2-Y^2}$ band 
and the bottom of the unfilled bands (the $M$ point) at 1 eV.

\begin{figure}[tbp]
{\resizebox{8cm}{5cm}{\includegraphics{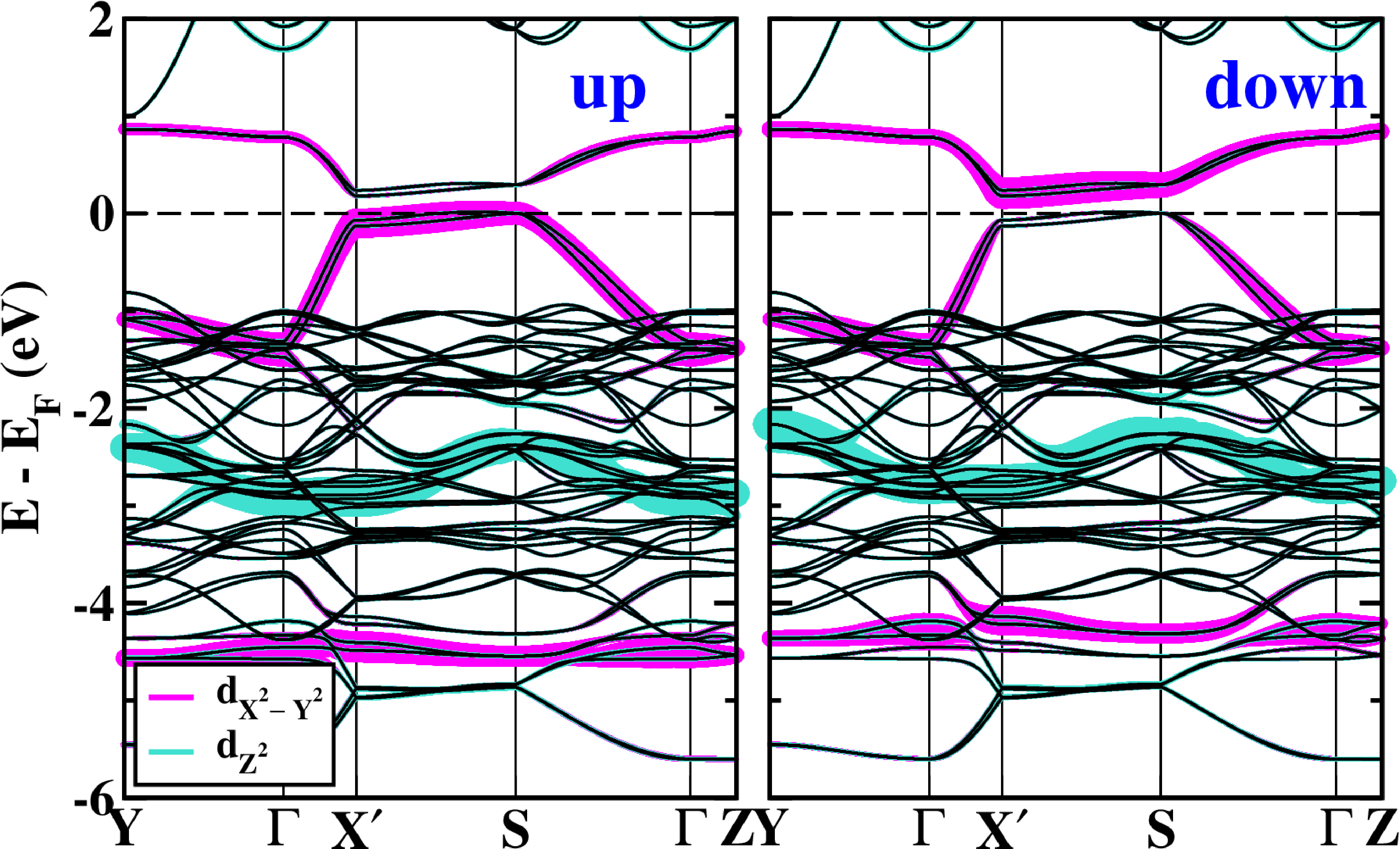}}}
\caption{AFM band structure of optimized \bco~ within GGA,  
highlighting the spin-resolved characters by fatband plots 
of the Cu $d_{X^2-Y^2}$ and $d_{Z^2}$ characters with a positive moment.
The AFM-induced gap indicates insulating character.
The $S$ point is the zone boundary in the (110) direction, 
as described in Fig. \ref{fs_sc}.
} 
\label{afmband}
\end{figure}

\begin{figure}[tbp]
{\resizebox{8cm}{6cm}{\includegraphics{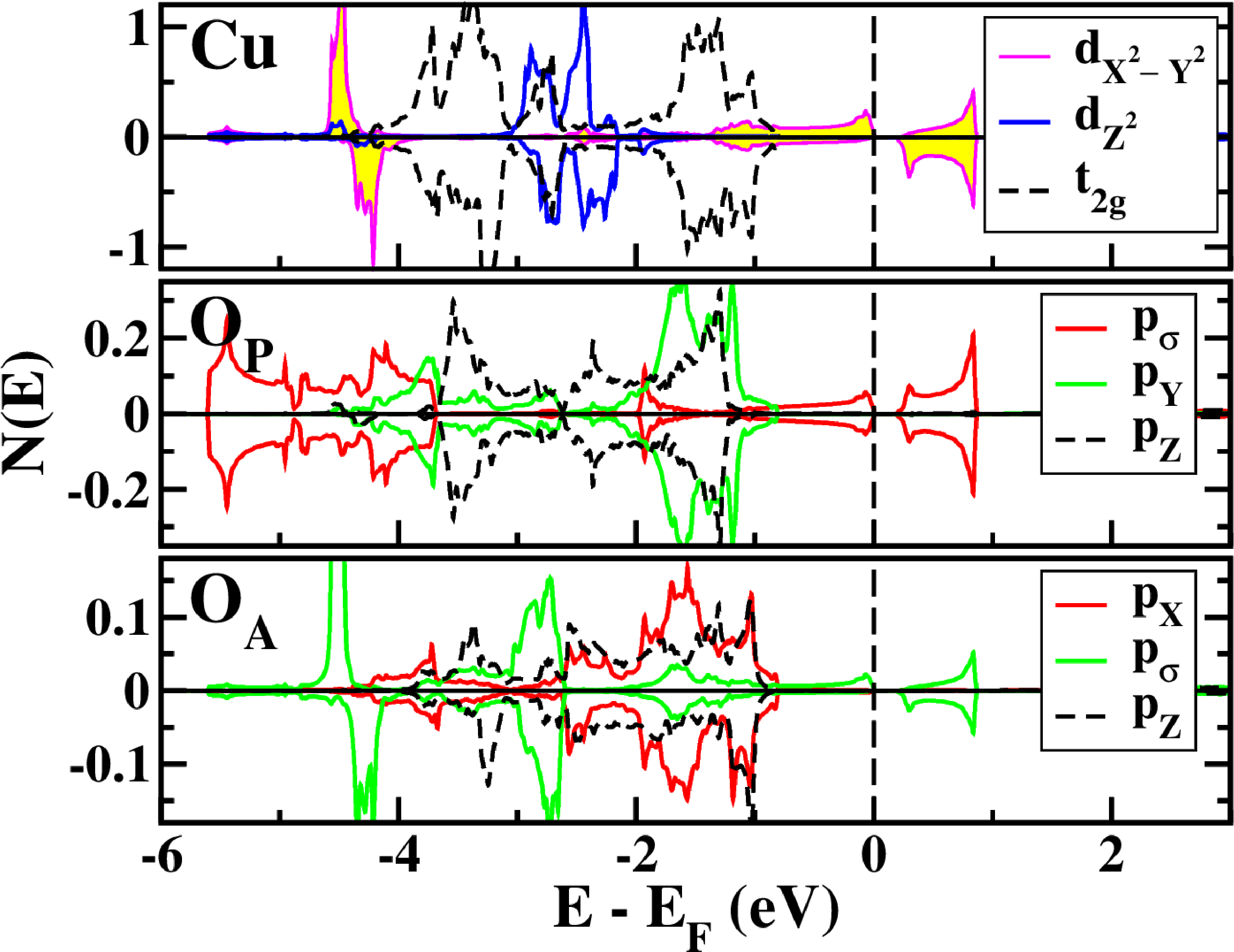}}}
\caption{GGA orbital-PDOSs of Cu, planar and apical oxygens in AFM  \bco.
The Cu ion with a positive moment is chosen.
Due to AFM symmetry, the only spin-polarized holes in the oxygens 
are on O$_{\rm A}$.
} 
\label{afmdos}
\end{figure}

\subsection{Antiferromagnetic state: Spin-Peierls instability}
The AFM state, which is the expected and calculated ground state of \bco~\cite{ladder2019}, 
was investigated using a supercell doubled  along the Cu-O$_{\rm P}$ chain direction.
The band structure in GGA is displayed in Fig. \ref{afmband}, showing 
an AFM-induced gap \cite{gap} of 0.2 eV.
The half-filled $d_{X^2-Y^2}$ band lies in the range of --1.5 eV to 1 eV,
whereas the fully-filled $d_{Z^2}$ orbital lies around --2.5 eV to --3 eV.
The corresponding orbital-PDOSs are given in Fig. \ref{afmdos}, 
indicating strong $pd\sigma$ hybridization, 
consistent with the spin density plot of Fig. \ref{str}(b).
The Cu spin moment, ideally a spin-half value of 1 $\mu_B$, is strongly
reduced to 0.18 $\mu_B$ by the $pd\sigma$ hybridization.
The O$_{\rm P}$ ions are thereby strongly polarized, but have zero 
net moment due to AFM symmetry.

A noteworthy aspect in the AFM state is a gap of 0.2 eV opening  
at the GGA level without the assistance of a Hubbard $U$ repulsion, which is
required in square-lattice cuprates. Gap opening without $U$
eliminates the categorization as a Mott insulator, leaving the
characterization of \bco~as in the Slater insulator regime, as
far as gap opening and the origin of magnetism are concerned. 
The zone-folding due to doubling of the cell along the 
$\hat{a}^\ast$-direction ($a^\ast=2a$)
leads to overlapped bands along the $X'-S$ line, {\it i.e.} 
on the $k_x=\pi/a^\ast$ plane. 
Then nesting of the 1D FS, evident in Fig. \ref{nmband}(b)
that it is near perfect, 
leads to a 1D spin-Peierls instability, analogous to the 
lattice-Peierls instability -- symmetry-breaking along the chain
is the underlying mechanism in both cases. 
As shown in the band structure of Fig. \ref{afmband},
along the $X'-S$ line the $d_{X^2-Y^2}$ bands bordering the gap have 
opposite spin character.
That is not the complete story. As in other cuprates, the
Hubbard $U$ surely has a strong effect, increasing the gap and moment
in the insulating phase and impacting spectral density spectra in the
metallic regime.

Inclusion of the Hubbard $U$ to the Cu ions simply 
increases the energy gap and the magnetic moment,
since the gap is already open in the GGA level.
At $U=7$ eV, the moment of the Cu ion reaches 0.64 $\mu_B$ and the gap is about 1.5 eV.

\subsection{Effects of oxygen mode displacements}
We have performed frozen phonon calculations for a few $\vec Q=0$ O displacements,
two involving O$_A$ motion and  one involving O$_P$ motion.
One is the beating $A_g$ mode of O$_{\rm A}$  ions against the Cu-O$_{\rm P}$ chain 
with amplitudes up to $\pm$0.06 \AA. 
The change in energy versus displacement (not shown here) 
is fit well with a simple harmonic form,
with frequency of 603 cm$^{-1}$, which is
within 7\% of the observed Raman frequency \cite{cwchu1994} of 563 cm$^{-1}$ 
in a multiphase sample of \bcox~with $\delta$ in the 0.09-0.17 range.

We have studied changes with an increase of the displacement in the IR active $B_{1u}$ mode
of the O$_A$-Cu-O$_A$ unit.
The energy gap linearly reduces with O displacement,  reaching 
$\Delta E_g$= --0.1 eV at the amplitude of $\pm0.06$ \AA~
that is a typical rms displacement of an oxygen ion (see SM \cite{sm}).
The frequency of the $B_{1u}$ mode is 560 cm$^{-1}$.

The O$_{\rm P}$ breathing mode along the chain direction was studied for
amplitudes $u_{\rm O}$ up to 0.07 \AA.
This broken symmetry results on two Cu sites alternating along the chain,
but the allowed spin disproportion is negligible, {\it i.e.}
spin-lattice coupling is small.
This oxygen optic mode has an anomalously low energy frequency of roughly 
43 cm$^{-1}$,  probably a remnant of the Peierls instability.
The electron-lattice coupling shows up in the modulation of an energy gap,
with  $dE_g/du_{\rm O}\approx$3.0 eV/\AA. 
The gap vanishes at 0.07 \AA; since this is a
typical oxygen rms vibrational amplitude, electron-phonon coupling in \bco~
may be worth further study.

\section{Results for the superconducting phase}
Before considering structures of ordered planar oxygens in the 
superconducting phase around $\delta\approx0.2$,
we carried out calculations of the virtual crystal approximation 
(VCA) up to $\delta=0.25$
by increasing the number of valence electrons from the $\delta=0$ phase.
As might be expected from the AFM DOS given in Fig. \ref{afmdos} 
and the FS $2k_F=\pi/a$,
once hole-doped the system moves away from the spin-Peierls 
instability and the gap and the small moment vanish.
In this region, the single-band feature at $\delta=0$ 
remains unchanged from the NM character (not shown here).
Above $\delta\approx0.25$, a small AFM moment re-appears as the
increased hole density begins to favor polarization.

\begin{figure}[tbp]
{\resizebox{8cm}{6cm}{\includegraphics{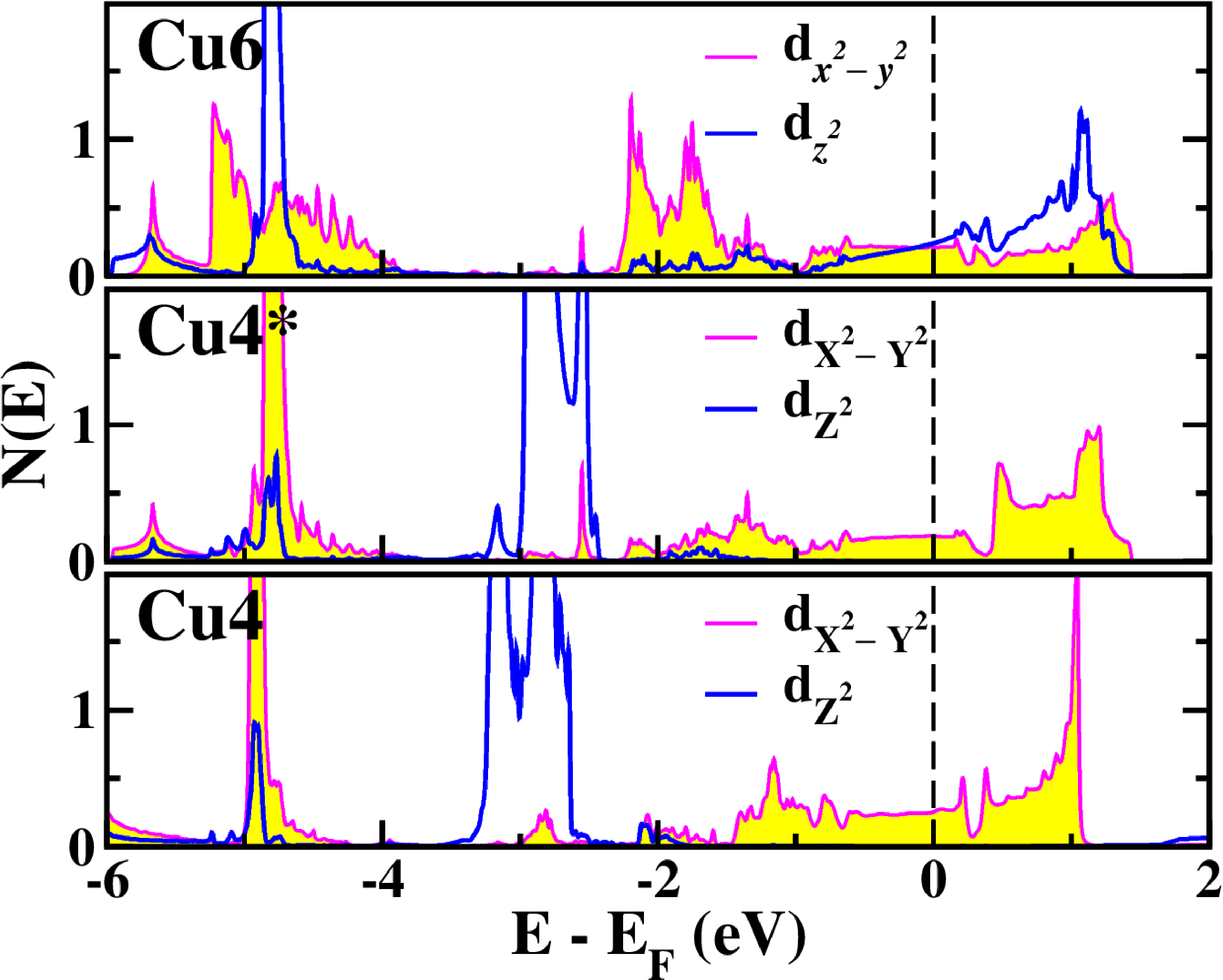}}}
\caption{ The Cu $e_g$-orbitals' projected DOSs for the NM $\delta=1/4$  
bilayer structure pictured in Fig. \ref{str_sc}(a), within GGA.
For the differing types of the local coordination, see the text.
} 
\label{pdos_sc}
\end{figure}

\subsection{Supercells}
We now turn to the supercell results at $\delta=1/4$.
The bilayer structure is favored,
with the monolayer cell being 70 meV higher and the
brickwall structure being 250 meV higher than the bilayer \cite{energy}.
These are of course ground state
(zero temperature) values. For varying high pressure and high temperature
synthesis conditions,
samples might contain regions with the metastable monolayer structure interspersed
with a majority amount of bilayer domains.
(Our band structure of the monolayer structure, shown in SM \cite{sm}, 
is similar to that published previously \cite{ladder2020}.)

Regularities across these structures based on coordination can be identified. The
PDOS spectra of the various Cu sites are provided in Fig.~\ref{pdos_sc} and the Appendix. 
In all three structures,\\
$\bullet$ the \underline{four-fold coordinated Cu4} chain 
sites display filled and strongly bound $Z^2$ orbitals while
the $X^2-Y^2$ bands are roughly half-filled.
This filling is
characteristic of a Cu$^{2+}$ ion with large crystal field splitting. 
Isolated chains in each structure
are likely to have strong AFM correlations, which (in the
absence of magnetic ordering, which is not yet reported) will reduce but not
eliminate conduction along the chain. See Sec. V for further description.\\
$\bullet$ the \underline{five-fold Cu5} sites in the two disfavored structures 
(see the Appendix) show 
holes distributed between both $e_g$ orbitals, reflecting the vast reduction
in the $e_g$ crystal field splitting compared to Cu4 sites. 
In the brickwall structure, the hole count 
is  dominated by $x^2-y^2$ weight ($z^2$ weight is minor); 
in the monolayer structure, surprisingly, $z^2$ hole weight dominates.
In each case the 
amount of Cu5 holes seems roughly characteristic of a formal valence of 
2.5+. \\
$\bullet$ for the \underline{four-fold rung Cu4$^{\ast}$} site, 
the hole(s) lies entirely in the
roughly half-filled $X^2-Y^2$ band, thus with one hole of each spin,
again close to a nominally Cu$^{2+}$ ion, as for the chain Cu4 ion. 
In the monolayer structure (see the Appendix), the Cu4$^{\ast}$ DOS
is gapped and clearly 2+.  Differing somewhat
from the brickwall structure, both Cu4 and Cu4$^{\ast}$ ions 
in the bilayer case have strongly
bound $Z^2$ states, with metallic but different from half-filled $X^2-Y^2$ 
bands. \\
$\bullet$ with two (chain layer) Cu4$^{2+}$ ions in the bilayer primitive cell, 
the average formal valence of the Cu6 and Cu4$^{\ast}$ ions must be Cu$^{2.5+}$.
The PDOSs in Fig.~\ref{pdos_sc} reveal that the \underline{six-fold Cu6} ion  
has {\it more $z^2$ holes than $x^2-y^2$ holes,} making it  
very highly oxidized (viz. approaching Cu$^{2.75+}$) and, given the overall highly
hole-doped character,  a likely candidate to be heavily involved
in superconducting pairing and in hosting the superconducting carriers.
Supporting this viewpoint is the expectation that strong spin correlations 
along the chain will reduce the participation of the Cu4 ion.
An interesting parallel is that $z^2$ holes have also been 
implicated \cite{bnoas,pickett2021} in the long sought and recently discovered 
superconductivity in infinite-layer hole-doped NdNiO$_2$ \cite{H.Hwang2019}.

\subsection{The Bilayer Oxygen Ordering}
We now focus on this energetically favored bilayer structure, 
whose electronic features have not been investigated previously.
In this phase, our calculations in {\sc wien2k} were performed 
with the experimental lattice and internal parameters given in
Ref.~[\onlinecite{pnas2019}].

In the bilayer structure of Fig. \ref{str_sc}(a),
the supercell, containing 4 formula units, 
has two Cu4 (chain) sites in the $z$=0 plane, and one planar Cu4$^{\ast}$
(rung) CuO$_4$ unit and one six-fold coordinated Cu6 in the $z=\frac{c}{2}$ plane.
In this structure, our attempts to obtain an AFM state (after all, in one layer
two well isolated chains remain intact) in GGA 
always reverted to a NM state. This result might be thought to be
consistent with the previous calculation \cite{ladder2019}, 
and our VCA results, if charge transfer between layers occurs.
Since the material is (super)conducting, we did not pursue possible magnetic
and insulating states using correlated electronic structure methods.

Figure \ref{pdos_sc} shows the $e_g$-orbitals' PDOSs for all three Cu sites.
Although the electronic structure is metallic, a formal valence 
viewpoint might still be instructive.
Since the Cu4 and Cu4$^{\ast}$ $Z^2$ states are both localized near --3 eV, 
their nominal valences (from $X^2-Y^2$ hole count) should be nearly the same.
On the contrary, the Cu6 sites contain hole(s) from both orbitals, 
with (surprisingly) somewhat more $z^2$ weight than $x^2-y^2$ weight. 
Figure~\ref{pdos_sc} indicates
that their crystal fields are drastically different: that of Cu6 is close
to an octahedral $e_g$ degeneracy.

\begin{figure}[tbp]
{\resizebox{7.5cm}{7.5cm}{\includegraphics{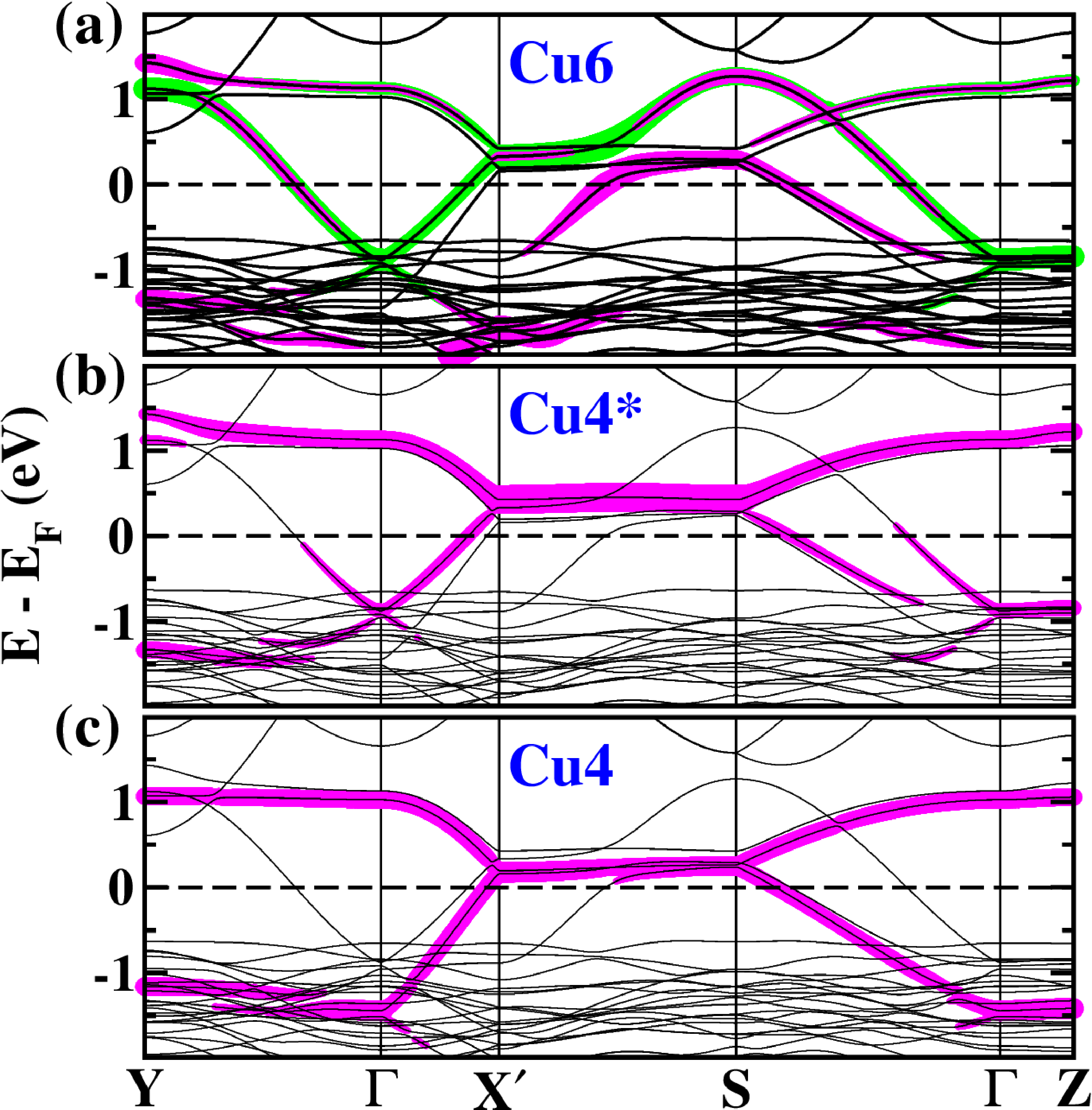}}}
\caption{ Band structure, enlarged in the range of --2 eV to 2 eV, 
of the NM $\delta=1/4$ bilayer phase, within GGA.
The characters of the $e_g$ orbitals of (a) Cu6, 
(b) Cu4$^\ast$, and (c) Cu4 are highlighted
by magenta ($d_{x^2-y^2}$ or $d_{X^2-Y^2}$) and green ($d_{z^2}$) colors.
}
\label{band_sc}
\end{figure}

\subsection{Charge, Crystal Fields, Bands}
{\it Charge density.} 
Bader charges can sometimes provide insight. They
do depend on environment and thus are not true atomic properties, and often
do not correspond closely to formal valence (atomic sphere charges also can vary
greatly from formal valences). Differences in Bader charge for inequivalent
ions give the most useful information.
Bader charges of the Cu ions are +1.31 (Cu6), +0.89 (Cu4$^\ast$), and +0.86 (Cu4).
The similarity of the latter two supports equal formal valence, and the 
difference of 0.42 to 0.45 relative to Cu6 indicates a 
substantially different valence for
Cu, again in agreement with the analysis above based on PDOSs.
The charges of the apical oxygens are $-1.12$ to $-1.28$, 
while those of the planar oxygens are  $-1.08$ to $-1.11$. These differences
of 0.03 to 0.2 likely reflect primarily differences in environment (bonding).
  

{\it Crystal fields.}
As expected from a simple crystal field concept, 
the site energy of the $d_{z^2}$ orbital (equal to that of $d_{x^2-y^2}$
for an ideal octahedron $e_g$ partner) changes 
with variation of the Cu-O$_{\rm A}$ distance $d_{\rm O_A}$.
We find that the $d_{z^2}$ orbital remains partially unfilled in the 
range of $d_{\rm O_A}$ distances from 1.86 \AA~ to 2.10 \AA.
The center of the $d_{z^2}$ band shifts by at most 0.1 eV in this range.
These results indicate that the effects of the shorter 
Cu-O$_{\rm A}$ distance on the electronic structure
are not substantial, consistent with the results of \bco.
In addition, as seen above in \bco, 
the corresponding variation of the total DOS around $E_F$ is negligible (not shown here).
This indifference to $d_{\rm O_A}$ suggests that this bondlength variation with doping
has little effect on $T_c$. 
At $E_F$ the total DOS is low and nearly identical to that of the NM $\delta=0$ phase, 
$N(E_F)=$0.73 eV$^{-1}$ per f.u. for both spins.

{\it Band structures.}
The corresponding band structure enlarged in the --2 eV to 2 eV 
region is shown in Fig. \ref{band_sc},
with the highlighted fatband plots of Cu6 $d_{x^2-y^2}$ and $d_{z^2}$ orbitals,
and Cu4$^\ast$ and Cu4 $d_{X^2-Y^2}$ orbitals. Note first the lack of
dispersion along $\Gamma$-$Z$ -- the layers remain uncoupled,
except for possible charge transfer.
The partially filled $d_{X^2-Y^2}$ orbitals of Cu4$^\ast$ and Cu4 ions
show very similar 1D dispersion, with widths of $\sim$2.4 eV, 
leading to $t_a\approx$0.6 eV.
The Cu4$^\ast$ orbital shows  
$d$-$d$ hybridization with Cu6 $e_g$ orbitals near $E_F$
along the $S$-$\Gamma$ line, due to accidental degeneracy and some small coupling.
On the other hand, the Cu4 orbital is negligibly hybridized 
with the other $d$ orbitals of Cu6 and Cu4$^\ast$, 
reflecting the 1D character of the chain.

Both (anti-bonding) Cu6 $d_{x^2-y^2}$ and $d_{z^2}$ 
orbitals of width 2.3 eV 
cross $E_F$  and are energetically nearly degenerate.
These Cu6 bands show two-dimensional (2D) character with $t_a\approx t_b\approx$0.56 eV.
As a result, within a formal charge picture, the Cu4 and Cu4$^\ast$ ions 
are Cu$^{2+}$ with half-filled $d_{X^2-Y^2}^1$ bands.
As noted, for formal charge balance, the Cu6 ion must be over-doped Cu$^{2.5+}$. 
Formal charges can be questionable for metallic systems, of course,
but the consistency of identification is striking.
This difference in Cu sites indicates that the ordered planar oxygen 
leads to a mixture of single-band and two-band features
that could significantly complicate determination of the superconducting pairing symmetry
and  mechanism.

\begin{figure}[tb]
{\resizebox{6cm}{3cm}{\includegraphics{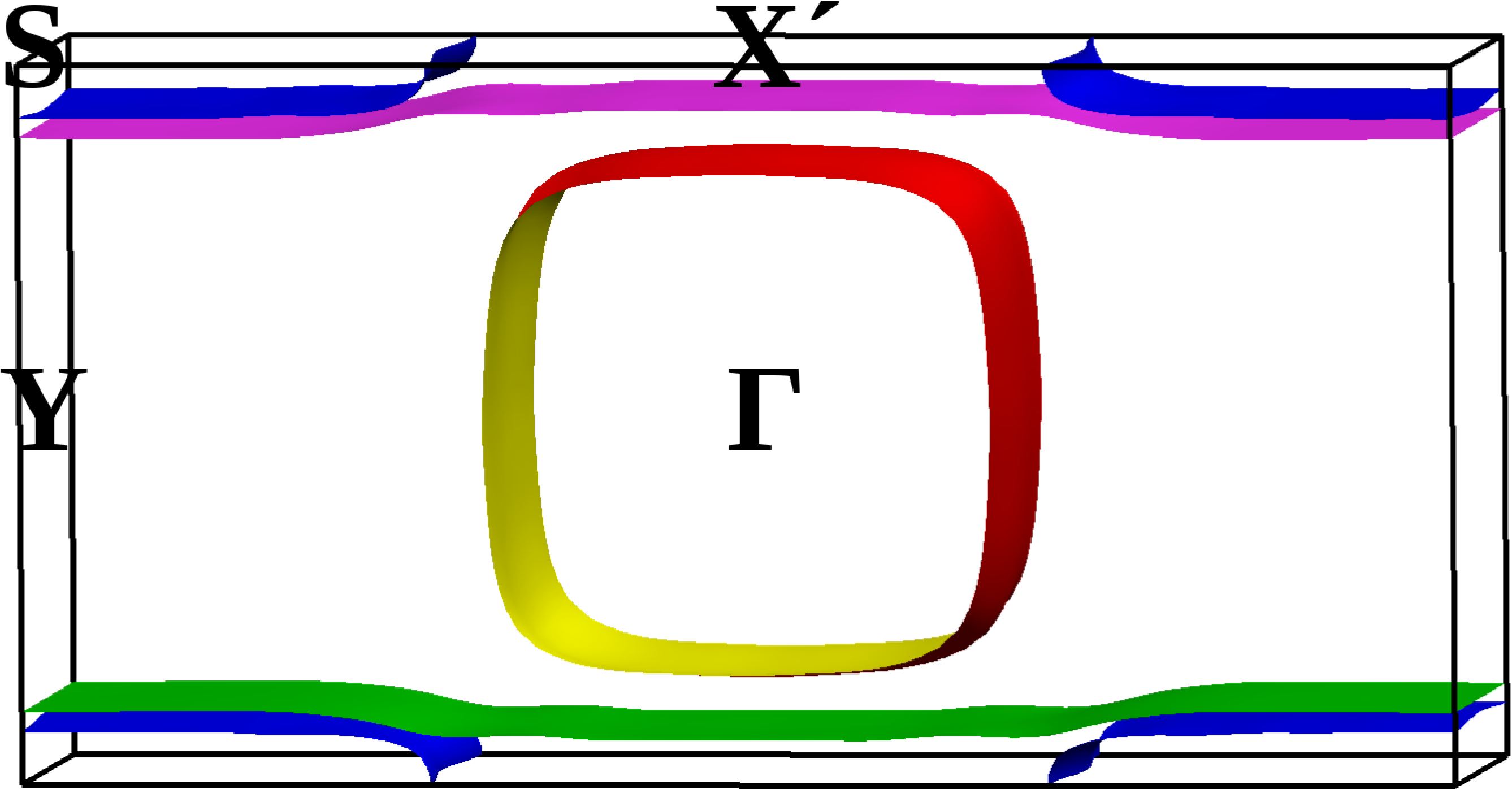}}}
\caption{Fermi surfaces within GGA of the NM $\delta=1/4$ phase 
in the bilayer structure, with several flat parallel sheets tending to
support charge, spin, or lattice instabilities.
} 
\label{fs_sc}
\end{figure}

\section{Fermiology of the bilayer structure}
Figure \ref{fs_sc} shows the Fermi surfaces of the NM bilayer $\delta=1/4$ phase.
All FSs show strongly reduced dimensionality.
They contain a large rounded-square barrel surface centered at $\Gamma$, 
with negligible dispersion 
in the third direction, analogous to several doped cuprate 
compounds. In  addition there are quasi-1D sheets near the $X'-S$ zone
edges, one extending along the entire edge length, the other cut off
midway by small in-plane dispersion. These surfaces are indicative of
strong 2D and 1D characters in the respective bands. 
The character of the barrel FS is a mixture of the Cu6 $e_g$ orbitals
with the Cu4$^\ast$ $d_{X^2-Y^2}$ orbital. 
The 1D FSs  have  mostly chain Cu4 and Cu4$^\ast$ $d_{X^2-Y^2}$ character, 
with minor admixture of Cu6 $d_{x^2-y^2}$.

The rms Fermi velocities over all sheets are 
$$v_{F,a}=24.7,~~ 
  v_{F,b}=4.5,~~
  v_{F,c}=0.05$$
 in units of 10$^7$ cm/sec,
a factor of 5 in planar anisotropy and likely at most hopping conductivity
in the perpendicular direction. These large velocities account
for the rather small Fermi level DOS in spite of large FSs.
This strong anisotropy is also reflected in the Drude plasma energies (in eV) of 
$$\Omega_{a}=3.9,~~  
  \Omega_{b}=1.66,~~  
  \Omega_{c}=0.17,$$
but smaller in magnitude.
In contrast to the two 1D FSs containing hole carriers with 
$v_{F,a}^{1D}\approx 34.5\times10^7$ cm/sec,
the $\Gamma$-centered barrel FS is less anisotropic in-plane:
$$v_{F,a}^b=7.2,~~
  v_{F,b}^b=11.1,~~ 
  v_{F,c}^b=0.08$$
 in units of 10$^7$ cm/sec.
This corner-rounded square FS with a side of 0.35$\frac{\pi}{a}$
possesses roughly 0.50 electrons, while hole carriers near the zone
boundary are contained within the 1D FSs.

One can note that this superconductor has a low $N(E_F)$ and 
low carrier density, a characteristic shared with other superconducting cuprates.
This is however different from the superconducting Fe-pnictides with their
higher $N(E_F)$ together with low carrier density \cite{singh2008}.
However, several superconducting cuprates show (within GGA) low $N(E_F)$ along with 
higher carrier density \cite{wep1992}.

\section{Discussion}
Bonner and Fisher demonstrated that quantum fluctuations preclude magnetic ordering
of the (isolated) spin-half Heisenberg chain at finite temperature \cite{bonner1964}, 
although strong spin-spin correlations lead to a vanishing susceptibility as
zero temperature is approached. Nevertheless, in the presence of even very small
interchain coupling, magnetic ordering is observed to occur in several
spin-chain materials. We have quantified the interchain coupling in Sec. III.
 
Schulz treated these intra-chain correlations exactly and interchain coupling
in mean field, and provided a formula for the staggered magnetization 
$m_0=\langle S^z_i \rangle$ at the $i$-th site and for the  N\'eel temperature $T_N$
of the 1D Heisenberg AFM spin-half chain \cite{schulz1996}.
For \bco, the staggered magnetization is
\begin{eqnarray} 
m_0\approx g\mu_B\times1.02\frac{t_b}{t_a}\approx0.17\mu_B,
\end{eqnarray}
with $g$ being the Land\'e $g$-factor. This result indicates the degree to
which 1 $\mu_B$ is strongly reduced by quantum fluctuations. Our calculations indicate
that  Cu-O $pd\sigma$ hybridization reduces the {\it static} ordered moment from 0.2 $\mu_B$
to 0.6 $\mu_B$ depending on the size of Hubbard $U$, which will be further
reduced by fluctuations. The magnetic susceptibility should contain signatures
of the strength of spin correlations as well as magnetic ordering.

Schulz also provided an implicit equation 
for ordering temperature $T_N$ in terms of simple interchain coupling $J_{\perp}$
and two constants $A$=0.32, $\Lambda$=5.8 from numerical calculations:
\beq
|J_{\perp}|=\frac{T_N}{4A\sqrt{{\rm ln}(\Lambda J/T_N)}}.
\eeq
With our estimates from Sec. III $J$=1640 K, $J_{\perp}\sim 10$ K, 
this expression gives $T_N$ in the neighborhood of 34K.
His expression with a single interchain coupling does not fit closely 
with the \bco~structure with several (small) interchain couplings that
we obtain, so we have used the largest. Within the 
uncertainty of the exchange couplings and the logarithmic dependence
of the expression, this is similar to but somewhat larger than the case
of Sr$_2$CuO$_3$ \cite{serge2020}, which means that  \bco~ is also
 near the regime of the 1D Luttinger-liquid quantum-critical phase.\\

In the superconducting sample \cite{pnas2019}, the linear specific 
heat coefficient $\gamma\sim14$ mJ/mol K$^2$
was experimentally measured.
Comparison of this value to our calculated band value of 
$\gamma_0=$1.72 mJ/mol K$^2$
leads to  a factor of 8 due to dynamic correlation effects, implying a 
strong correlation strength in the doped material.

Our calculations indicate that this system is little affected 
by the unusual Cu-O separations.
This behavior can be contrasted with the infinite layer Ni$^{2+}$ 
system of Ba$_2$NiO$_2$(AgSe)$_2$, 
in which the strained Ni-O separation leads to  a calculated
`off-diagonal singlet' state \cite{bnoas},
consistent with previous findings by some of the current authors 
that Cu ions are different from Ni ions even if both have the same 
ionic configuration \cite{lanio2,nno2}.

The bilayer FSs suggest instabilities of the ordered structure that may
help in understanding the paucity of single phase, well ordered samples.
Flat (or parallel -- nesting) regions of FSs provide a strong tendency
toward electronic instability that manifests as charge, spin, or
lattice instabilities, each of which break the symmetry of the underlying phase. 
The barrel FS shows nesting at $2k_F^b\sim 0.70\frac{\pi}{a}$ in both of the
planar directions.  The 1D FSs suggest instabilities at a small value of
$2k_F^{1D}$ in the $\hat{a}$-direction ({\it i.e.} a long wavelength disturbance),
and a large value $2k_F^{1D}$=$\frac{2\pi}{a} - 2k_F^{1D}$, which corresponds to a
short wavelength disturbance. These potential or real instabilities may 
help to account for the complex structures observed in samples. 

\section{Summary}
Our first principles studies of the electronic and magnetic
structures of \bcox, $\delta=0$ or $\frac{1}{4}$, provide important information
about this unusual superconductor. Our studies indicate that interchain 
coupling in NM (hence metallic) \bco~is quite small, thus not surprisingly,
this state subject to Peierls-type instability of spin, of lattice,
and possibly also of charge ordering. Considering other cuprates, we have
focused on the spin (AFM) instability. AFM ordering alleviates the
Peierls instability, however the magnetism is weak without the assistance of
intra-atomic Cu $3d$ repulsion and correlation. Charge and lattice instabilities
would compete with AFM ordering. If AFM ordering occurs, it will be weak
due to the fact that interchain coupling is very small, and isolated chains
do not order -- the Bonner-Fisher result.
Considering the very stable Cu$^{2+}$ ion it is surprising that
stoichiometric, ordered \bco~ has not been obtained, even in samples synthesized
with varying pressure and temperature.

Isolated 1D spin-half chains do not order due to strong fluctuations, however
even small interchain coupling does lead to order. The large intrachain and
small interchain couplings that we estimate, using the theory of
Schulz, suggest an ordering temperature of the order of 34 K. This value
puts \bco~in the regime of the 1D Luttinger-liquid quantum critical phase,
for which further experimental study will be useful. 

The types of ordering of the planar oxygen dopants play a 
crucial role in determining the crystal and electronic structures of \bcox, and thereby
the value of $T_c$ and pairing mechanism.
This observation is consistent with experimental data that show 
substantial change of $T_c$ 
related to the modulation of apical oxygen positions and concentration
in superconducting \scox~\cite{liu2014,pnas2020}. The doping level of
$\delta$=1/4 has led to the picture of ordered approximants being modeled
as additional O ions being inserted, seemingly necessarily bridging Cu-O$_{\rm P}$ chains,
rather than removing a much larger fraction of O ions from Ba$_2$CuO$_4$.

Our study of three proposed ordered structures of \bcox~has indicated that a
{\it bilayer} structure (one with two structurally different layers) is 
most stable, although a {\it monolayer} ordered structure
is only modestly higher in energy. In the bilayer structure, all three Cu
ions have $e_g$ states at the Fermi energy and participate in conductivity,
hence presumably also in superconductivity. There are however strong site
 differences: the two four-fold coordinated sites have the $z^2$ orbital
strongly bound and localized, leaving only more or less half-filled
$x^2-y^2$ holes as occurs in most cuprates. 

The octahedrally coordinated site, which occurs only in
the bilayer structure, has a large fraction of holes in both $e_g$ orbitals,
with more in the $z^2$ orbital. This marks this site as very highly
oxidized (viz. 2.75+), accounting for the perplexing high mean formal
valence of \bcox.  A quasi-degeneracy of the $e_g$ orbitals is restored
by the octahedral coordination, and implies that \bcox~ is not only strongly
overdoped but also a two-band superconductor. 

In addition, this energetically preferred bilayer structure 
displays intense Fermi surface nesting  of both 1D and 2D nature that will 
drive the system toward instabilities of a few types (wavelengths) of 
charge-, spin-, and lattice-symmetry breaking. The fluctuation associated
with these instabilities may help to account for \bcox~not having been 
prepared in a fully ordered, single phase sample even for a variety of
synthesis conditions.  

The current work on the metallic phase has not dealt 
specifically with correlation effects due to
strong repulsion in the $3d$ shell of the Cu ion. Crystal fields separating
the two $e_g$ orbitals on the four-fold and six-fold coordinated Cu sites have
been identified and analyzed, and formal valence identifications have been 
discussed. Specifically, four-fold sites qualify as 2+, with the others 
better characterized as 2.5+ or even somewhat more highly oxidized. 
Our results provide a roadmap for
further analysis and experimental exploration of the electronic structure of \bcox.

{\it Note added.} Recently, a dynamical mean field work similar to this appeared in Ref. \cite{held}.
In their study they used $U$=3 eV, but for the superconducting phase,
while we treat only the insulating phase with DFT+U.

\section{Acknowledgments}
H.S.J. and K.W.L. were supported by National Research Foundation of Korea
Grant No. NRF-2019R1A2C1009588.
W.E.P. was supported by NSF Grant No. DMR 1607139.

\begin{figure*}[tbp]
{\resizebox{8cm}{6cm}{\includegraphics{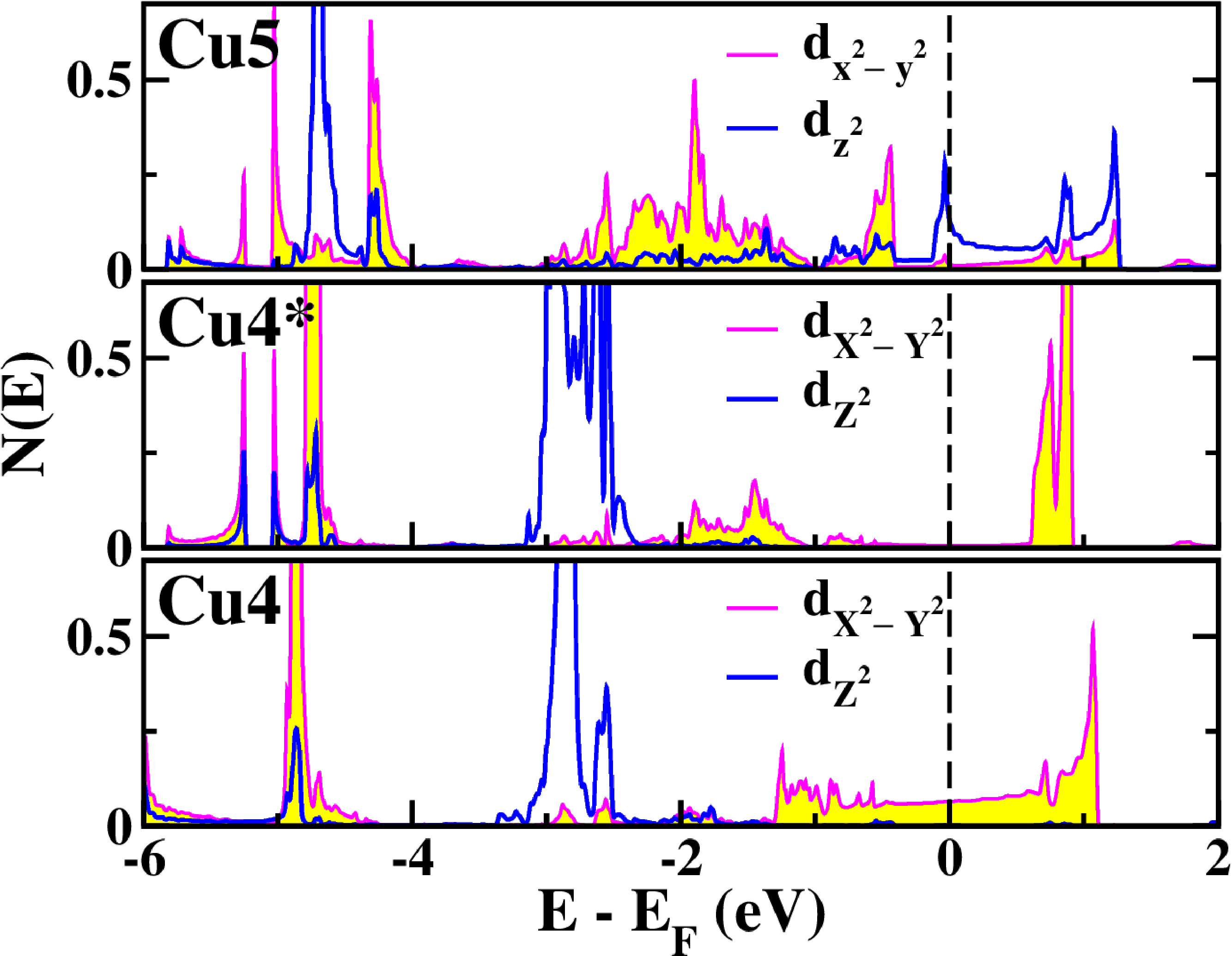}}}
\hskip 4mm
{\resizebox{8cm}{6cm}{\includegraphics{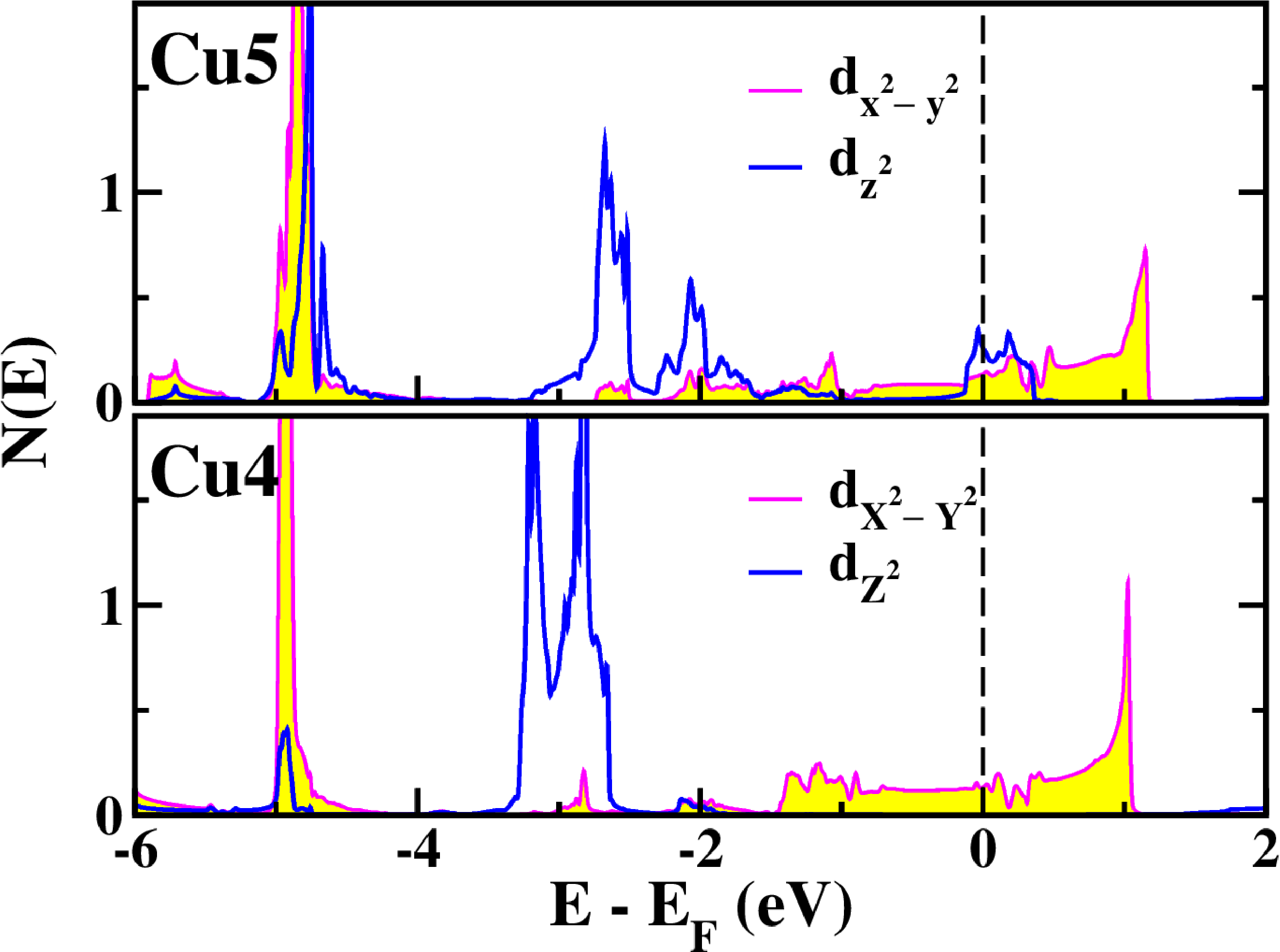}}}
\caption{ The Cu $e_g$-orbitals' PDOSs for the NM $\delta=1/4$  
(left panel) monolayer structure pictured in Fig. \ref{str_sc}(b)
and (right panel)  brickwall structure pictured in Fig. \ref{str_sc}(c), within GGA.
} 
\label{appendix}
\end{figure*}

\section{Appendix}
Since the PDOSs and associated crystal field splitting of the Cu $e_g$ orbitals
play a strong role in our identification of the Cu sites and orbital character that
provide the superconducting carriers, we provide here, for 
the $\delta=1/4$ phase, the Cu $e_g$-orbitals' PDOSs 
for the energetically unfavored structures in Fig. \ref{appendix}.
Strong differences can be seen in the $d_{z^2}$ spectra: for the Cu5 site,
there is strong character at and above $E_F$, while that orbital is localized
at --3 eV in the two four-fold sites, 
while great similarities are evident. 
It is not surprising that the ``isolated'' chain site Cu4 spectrum is relatively
undisturbed by doping on other layers as in \bco. There is however a strong
difference between Cu4 and Cu4$^{\ast}$: the former is half-filled, while
the latter is gapped around $E_F$, seemingly approaching a 3+ formal valence.
The brickwall Cu5 site (the right panel of Fig. \ref{appendix}) 
likewise approaches this same high formal valence. The main text provides 
additional discussion and comparison of all three structures.



\begin{thebibliography}{10}
\bibitem{optimize}
 In our structure optimization of \bco,
 the $a=b$ case shows similar Cu-O$_{\rm A}$ distance and 
 smaller  Cu-O$_{\rm P}$ distance  by 0.06 \AA, compared with the $a\not=b$ case.
\bibitem{energy} 
Structural relaxation was carried out with the {\sc vasp} code.
The energy differences we quote are for that structure, but
calculated with the {\sc wien2k} code. The energy differences are
somewhat different from the two codes but semi-quantitatively in
agreement.
\end{thebibliography}
\end{document}